\documentclass[%
reprint,
amsmath,amssymb,
aps,
showkeys
]{revtex4-2}

\usepackage[braket, qm]{qcircuit}
\usepackage{graphicx}
\usepackage{dcolumn}
\usepackage{bm}
\usepackage{xcolor}
\usepackage{hyperref}
\hypersetup{colorlinks,
allcolors=violet}
\usepackage{bbm}
\usepackage{soul}
\usepackage{qcircuit}
\usepackage{amsmath}
\usepackage{float}
\usepackage{multirow}
\usepackage{longtable}
\usepackage{booktabs}
\usepackage{adjustbox}
\usepackage{longtable}
\usepackage{siunitx}
\usepackage{array}

\begin{document}

\title{Practical Fidelity Limits of Toffoli Gates in Superconducting Quantum Processors}
\author{M. AbuGhanem{$^{1,2}$}}
\address{$^{1}$ Faculty of Science, Ain Shams University, Cairo, $11566$, Egypt}
\address{$^{2}$ Zewail City of Science, Technology and Innovation, Giza, $12678$, Egypt}
\email{gaa1nem@gmail.com}

\date{May 9, 2025}

\begin{abstract}

High-fidelity multi-qubit gates are a critical resource for near-term quantum computing, as they underpin the execution of both quantum algorithms and fault-tolerant protocols. The Toffoli gate (CCNOT), in particular, plays a central role in quantum error correction and quantum arithmetic, yet its efficient implementation on current quantum hardware remains limited by noise and connectivity constraints. In this work, we present a hardware-aware characterization of the Toffoli gate using optimized, connectivity-compliant decompositions executed on IBM’s 127-qubit superconducting quantum processors. Our study integrates state preparation, gate synthesis, and quantum state/process tomography (QST/QPT) to evaluate fidelity across three distinct classes of input states: Greenberger–Horne–Zeilinger (GHZ), W, and the uniform superposition of all three-qubit computational basis states—under noise-free simulation, noise-aware emulation, and real hardware execution. For GHZ states, we report state fidelities of 98.442\% (noise-free simulation), 81.470\% (noise-aware quantum emulation), and 56.368\% (real quantum hardware). For W states, state fidelities are 98.739\%, 79.900\%, and 63.689\%, respectively, and for the uniform superposition state, we observe state fidelities of 99.490\%, 85.469\%, and 61.161\%. Comparative QPT experiments yield process fidelities of 98.976\% (noise-free) and 80.160\% (noise-aware emulation). Our results empirically characterize state-dependent error patterns in multi-qubit circuits and quantify trade-offs between gate decomposition strategies and native hardware performance, offering practical insights for scalable, hardware-efficient quantum circuit design.

\end{abstract}


\maketitle

\section{Introduction}

Quantum computers~\citep{qc,IBMQuantum} harness the principles of quantum superposition and entanglement~\citep{Theprinciples,AbuGMSc19} to solve problems intractable for classical systems~\citep{art19,Zuchongzhi,Light,Zuchongzi2.1,Kim23,NISQ24,Willo25,GoogleQuantum}, from simulating quantum materials~\citep{Quantumchemistrysimulation} and drug design~\citep{Drugdesign,DrugDiscovery_review} to advancing global supply chains~\citep{supplychains} and accelerating cryptographic breakthroughs~\citep{Shor,factor2048bitRSA}. Quantum computers holds immense promise for real-world applications~\citep{IBMQuantum}, enabling industries to tackle complex challenges with unprecedented efficiency, accuracy, and cost-effectiveness~\citep{IBMQuantum,PhotonicQuantumComputers}. While fault-tolerant architectures—systems capable of maintaining computational accuracy despite inherent noise, decoherence, and hardware imperfections—promise transformative applications~\citep{Fault-tolerant,Fault-tolerant_anyons}. Today’s noisy intermediate-scale quantum (NISQ) processors~\citep{NISQ24} face stringent limitations in qubit count, coherence times, and gate fidelity~\citep{NISQ18}. Realizing practical advantages on these systems demands efficient implementations of fundamental operations—particularly multi-qubit gates that encode conditional logic~\citep{requirementstopracticalquantumadvantage}.

Among these, the Toffoli gate~\citep{Toffoli} stands as a cornerstone of quantum computation, enabling universal classical and quantum logic when combined with single-qubit operations~\citep{MikeIke}. Its ability to perform controlled operations on three qubits makes it indispensable for quantum algorithms ranging from error correction~\citep{Fault-tolerant,Fault-tolerant_anyons} to arithmetic circuits~\citep{1stFirstToffolirealization}. However, implementing high-fidelity Toffoli gates on NISQ processors remains a challenge, as the gate’s complexity exacerbates the effects of decoherence, crosstalk, and calibration errors~\citep{NISQ18}.

While superconducting quantum processors~\citep{SuperconductingQuantumComputers} have demonstrated remarkable progress in one- and two-qubit gate fidelities~\citep{Elementarygates,Kim23,NISQ24,Willo25}, three-qubit gates like the Toffoli gate~\citep{Toffoli} often require decomposition into native gates, introducing additional overhead and noise~\citep{Synthesis_circuits,Errormitigation}. Recent work has explored both direct implementations~\citep{MikeIke,Synthesis7} and optimized decompositions (as discussed in Section~\ref{SEC:Hardware-Aware}), but a systematic characterization of the gate’s performance across different input states—particularly entangled states relevant to algorithms—is still lacking.

In this work, we present a comprehensive experimental study of the Toffoli gate~\citep{Toffoli} performance on state-of-the-art superconducting processors from IBM Quantum (\textit{ibm\_sherbrooke} and \textit{ibm\_brisbane})~\citep{IBMQuantum}. To understanding their behavior across diverse input states on real hardware, using full quantum state tomography, we evaluate the gate’s state fidelity for three critical input states: Greenberger–Horne–Zeilinger (GHZ) states (\( (|000\rangle + |111\rangle)/\sqrt{2}\)), testing coherence in maximally entangled systems;  
W states (\((|001\rangle + |010\rangle + |100\rangle)/\sqrt{3}\)), sensitive to asymmetric errors;  
Uniform superpositions (\(\sum_{x \in \{0,1\}^3}|x\rangle /\sqrt{8}\)), probing basis-state-independent behavior.

Quantum tomography provides a powerful lens to evaluate gate performance beyond average fidelity metrics. QST reconstructs the output density matrix, revealing how specific input states degrade under noise~\citep{QPTGambetta}. Meanwhile, QPT fully characterizes the gate’s operation, identifying dominant error channels~\citep{qpt,chuang97}. These tools are essential for debugging and optimizing gates in practice~\citep{AGSycam1,AGSycam2,qpt_Mølmer,Entangling}.

We complement these measurements with a full QPT analysis of the Toffoli gate~\citep{Toffoli} on both noiseless quantum simulator and
noisy quantum emulator; by employing a comprehensive tomography framework that combines quantum QST and QPT to evaluate gate performance in three key regimes. First, we assess entanglement resilience using GHZ and W states, which serve as stringent tests of coherence under maximal entanglement conditions. Second, we examine basis-state uniformity through uniform superposition states, revealing systematic biases in the computational basis. Finally, by comparing real quantum hardware results with both noiseless quantum simulator and noisy quantum emulator benchmarks, we precisely quantify noise-induced deviations in gate operation. Our results help bridge the gap between theoretical gate specifications and real-world implementation fidelity by providing detailed benchmarks for three-qubit operations in NISQ-era quantum computing. The state-dependent error profiles and simulator-to-hardware comparisons we report provide a roadmap for optimizing three-qubit gates in NISQ-era quantum algorithms~\citep{NISQ-algorithms}.

The remainder of this work is structured as follows: 
Section~\ref{SEC:Hardware-Aware} reviews hardware-aware quantum circuit synthesis, with a focus on optimized Toffoli gate implementations for 
NISQ devices (Section~\ref{SEC:CompilationStrategies}). 
Section~\ref{SEC:ExistingImplementations} discusses hardware-specific decomposition techniques. 
Section~\ref{sec:Methods} details our experimental setup, including:
gate implementation protocols (Section~\ref{sec:Methods_Gate}),
state preparation methodologies (Section~\ref{sec:method_state}), and 
quantum tomography procedures for fidelity validation (Section~\ref{sec:method_tomography}). 
Section~\ref{SEC:results} analyzes results and discussion, comparing theoretical predictions with empirical data. 
Section~\ref{sec:QuantumHardwarePerformance} evaluates quantum hardware performance metrics across different superconducting quantum architectures. 
Lastly, Section~\ref{SEC:conclusion} concludes with key findings and future directions for scalable quantum circuit synthesis.

\section{Hardware-Aware Quantum Circuit Synthesis and Toffoli Gate Implementations}\label{SEC:Hardware-Aware}

\subsection{Compilation Strategies for Robust Implementations}\label{SEC:CompilationStrategies}

The optimization of quantum circuits is critical for enabling practical applications across dynamical simulations and combinatorial problem-solving. Current methodologies can be classified into several key approaches: 
problem-informed techniques that exploit dynamical symmetries~\citep{Nemkov_14,Nemkov_13,Nemkov_12,Nemkov_11}, 
noise-resilient compilation strategies~\citep{Nemkov_18,Nemkov_17,Nemkov_16,Nemkov_15}, 
hardware-adaptive gate set redefinitions~\citep{Nemkov_19}, 
and numerical circuit synthesis leveraging computational search algorithms~\citep{Nemkov}.

Recent advances in circuit synthesis have been propelled by enhanced computational resources and sophisticated optimization frameworks. These include discrete searches over finite gate sets~\citep{Nemkov_22, Nemkov_21}, 
hybrid discrete-continuous optimization~\citep{Nemkov_24, Nemkov_23}, 
adaptive synthesis protocols~\citep{Nemkov_28,Nemkov_27,Nemkov_26,Nemkov_25}, and 
metaheuristic methods such as 
genetic algorithms~\citep{Nemkov_30, Nemkov_29} and
machine learning~\citep{Nemkov_32, Nemkov_31}. 
Notably, emerging hybrid techniques~\citep{Nemkov_34, Nemkov_33} integrate architectural search with continuous parameter optimization, building upon principles from variational compilation of random unitaries~\citep{Nemkov_34,Nemkov}—a standard benchmark for assessing circuit complexity~\citep{Nemkov_37,Nemkov_36,Nemkov_35}.

To execute high-level quantum algorithms on physical hardware, high-level circuit descriptions must be translated into device-specific native operations. This compilation process typically employs quantum assembly languages such as OpenQASM~\citep{cross2021openqasm3} as intermediate representations. Although quantum instruction sets fundamentally comprise 1-qubit rotations and 2-qubit CNOT gates, actual hardware implementations frequently demand additional decomposition into architecture-specific native gate sets. For instance, fixed-frequency superconducting quantum processors utilize the echoed cross-resonance (ECR) gate as their fundamental two-qubit operation~\citep{ref10}.

The development of optimal gate decomposition methods represents a fundamental challenge in quantum compilation. Academic investigations have pursued two complementary directions: general techniques for arbitrary unitary operations~\citep{mottonen2004synthesis2,vartiainen2004synthesis, shende2006synthesis3,davis2020topologyaware} and specialized implementations for common quantum circuits, including Toffoli gates~\citep{jones2013lowToffoliTgate} and quantum Fourier transforms~\citep{cleve2000parallelQFT}. Modern approaches incorporate detailed hardware constraints into the compilation process, developing topology-aware decomposition methods that significantly reduce gate counts and circuit depth~\citep{cheng2018toffolilinear,hu2019mctsquare}.

Leading quantum compilation frameworks, including 
Qiskit~\citep{qiskit}, 
t$\ket{\text{ket}}$~\citep{sivarajah2020tket}, and
Cirq~\citep{cirq},
implement sophisticated gate-level optimization pipelines. 
These systems incorporate multiple complementary techniques: 
redundant gate elimination through cancellation methods~\citep{maslov2008gatecancellation}, 
optimal two-qubit gate synthesis via KAK decomposition~\citep{kraus2001kak1}, 
and adaptive circuit restructuring through dynamic gate reordering~\citep{liu2022NASSC}. 
Compilers additionally employ specialized optimization strategies such as 
localized peephole circuit transformations~\citep{peephole}, 
software-based crosstalk suppression~\citep{murali2020softwarecrosstalk}, 
and noise resilience through dynamical decoupling sequences~\citep{smith2022timestitch}. 
Additionally, The QContext compiler~\citep{QContext}, which pioneers a context-aware synthesis approach. This method simultaneously optimizes both abstract circuit representations and hardware-specific native gate implementations.

\subsection{Multi-Qubit Toffoli Gates}\label{SEC:M-TOFFOLI}

As a paradigmatic multi-qubit operation, the Toffoli gate~\citep{Toffoli} represents a critical building block in quantum information processing, combining multiple control qubits with a single target qubit to enable conditional state manipulation~\citep{MikeIke}. First introduced by Tommaso Toffoli~\citep{Toffoli}, this gate achieves computational universality when augmented with Hadamard operations~\citep{MikeIke,TOFFOLI_51,TOFFOLI_52}, while simultaneously serving as a fundamental construct in reversible computing paradigms that connect classical and quantum computational models~\citep{TOFFOLI_53}. 
The gate's operational versatility is demonstrated through its essential role in cornerstone quantum algorithms including 
unstructured search (Grover's algorithm)~\citep{MikeIke,GSA},
prime factorization (Shor's algorithm)~\citep{TOFFOLI_55,TOFFOLI_54}, 
and fault-tolerant circuit designs~\citep{TOFFOLI_57,TOFFOLI_58}, as well as
various quantum error correction schemes~\citep{TOFFOLI_59,TOFFOLI_60,TOFFOLI_61,GoogleQuantum}.

Formally, the $k$-qubit Toffoli gate family comprises $k-1$ control qubits governing the state inversion of a single target qubit, with the operation executing only when all control qubits occupy the $\ket{1}$ state~\citep{MikeIke}. The simplest non-trivial instance ($k=2$) reduces to the ubiquitous CNOT gate, while the 3-qubit variant (denoted as \(\hat{\mathcal{V}}_{\text{Toffoli}}\)) has emerged as particularly significant in quantum information theory.

This deceptively simple operation enables non-trivial computational power when combined with single-qubit gates, forming a universal gate set for quantum logic~\citep{Elementarygates}.  
The underlying unitary transformation and quantum circuit representation for this three-qubit operation is given by:
\begin{equation}\label{ccx_circuit_matrix}
\begin{aligned}
\hat{\mathcal{V}}_{\text{Toffoli}} &= 
\begin{array}{c}
\Qcircuit @C=0.7em @R=1em @!R {
    \lstick{} & \ctrl{2} & \qw \\
    \lstick{} & \ctrl{1} & \qw \\
    \lstick{} & \targ & \qw
}
\end{array}
=
\begin{pmatrix}
1 & 0 & 0 & 0 & 0 & 0 & 0 & 0 \\
0 & 1 & 0 & 0 & 0 & 0 & 0 & 0 \\
0 & 0 & 1 & 0 & 0 & 0 & 0 & 0 \\
0 & 0 & 0 & 1 & 0 & 0 & 0 & 0 \\
0 & 0 & 0 & 0 & 1 & 0 & 0 & 0 \\
0 & 0 & 0 & 0 & 0 & 1 & 0 & 0 \\
0 & 0 & 0 & 0 & 0 & 0 & 0 & 1 \\
0 & 0 & 0 & 0 & 0 & 0 & 1 & 0 \\
\end{pmatrix} \\
&\equiv 
\begin{pmatrix}
\mathbb{I}_4 & \mathbf{0}_4 \\
\mathbf{0}_4 & \hat{\mathcal{V}}_{\text{CNOT}}
\end{pmatrix}
\end{aligned}
\end{equation}

Beyond its quantum applications, the Toffoli gate's classical simulation capability is noteworthy~\citep{Toffoli}, as it can replicate any classical logic circuit while maintaining reversibility~\citep{Toffoli}. However, practical implementations face complexity barriers. Unlike native one- and two-qubit gates, three-qubit gates often require decomposition into lower-level operations—a process that introduces gate-depth overhead and amplifies errors~\citep{Synthesis_circuits}. For instance, typical implementations (assumes an ideal fully-connected quantum architecture) requiring decomposition into at least five 2-qubit gates~\citep{TOFFOLI_49} or six CNOT operations~\citep{MikeIke} when restricted to elementary 1- and 2-qubit gates~\citep{TOFFOLI_48}. These implementation challenges are further compounded in distributed quantum systems, highlighting the intricate trade-offs between gate complexity and operational fidelity in physical realizations.

\subsection{Hardware-Specific Decomposition Strategies}\label{SEC:ExistingImplementations}

\begin{figure*}[htp!]
    \centering
    \includegraphics[width=0.7\textwidth]{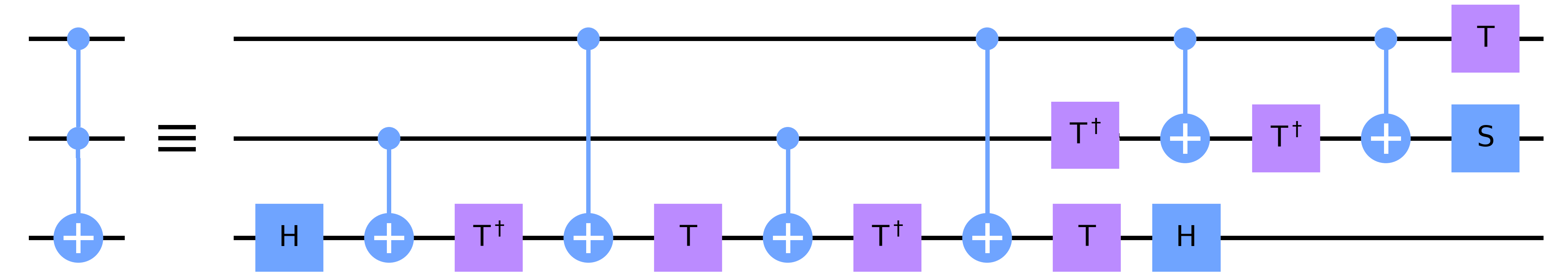}
    \caption{A standard decomposition of a three-qubit CCX (Toffoli) gate requires six CNOT gates along with multiple single-qubit gates~\citep{MikeIke}. This conventional approach assumes an ideal fully-connected quantum architecture, a condition rarely met in NISQ-era devices due to their limited connectivity. This decomposition uses the following single-qubit gates: 
\( 
 H = \frac{1}{\sqrt{2}} \begin{pmatrix} 1 & 1 \\ 1 & -1 \end{pmatrix}, \quad 
 T = \begin{pmatrix} 1 & 0 \\ 0 & e^{i\pi/4} \end{pmatrix}, \quad 
 S = \begin{pmatrix} 1 & 0 \\ 0 & i \end{pmatrix}.
\)
}
\label{fig:CCX_6_CX}
\end{figure*}

The Toffoli gate~\citep{Toffoli} has been implemented across various superconducting qubit architectures~\citep{SuperconductingQuantumComputers}, with the canonical linear decomposition remaining one of the most widely used configurations~\citep{MikeIke}. However, practical deployment on NISQ devices is complicated by limited qubit connectivity, necessitating decompositions with minimal gate overhead. For instance, a three-qubit CCX gate typically requires at least six CNOT gates alongside single-qubit operations~\citep{MikeIke}. Typical decompositions use sequences of CNOT and  single-qubit gates as depicted in Figure~\ref{fig:CCX_6_CX}, but these may not align optimally with a processor’s native gate set or connectivity.  

\begin{figure*}[htp!]
    \centering
      \includegraphics[width=0.7\textwidth]{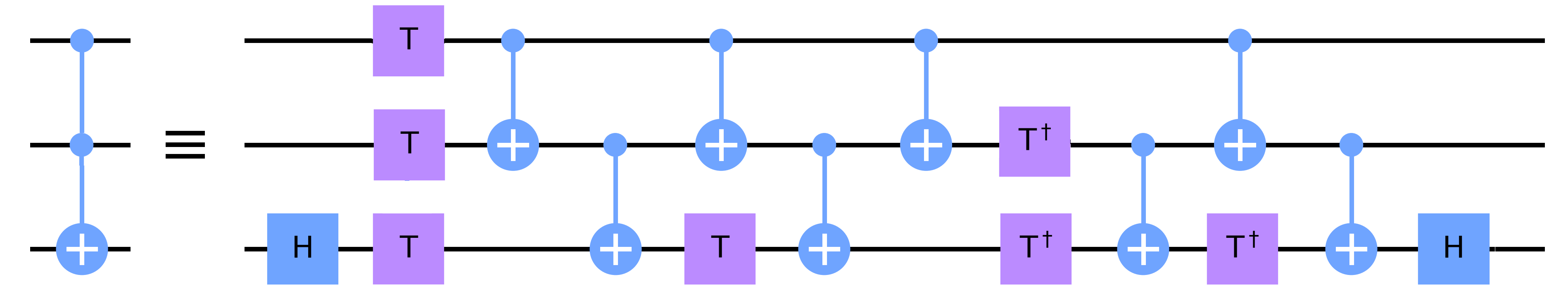}
    \caption{A decomposition of a three-qubit CCX (Toffoli) gate requires 8 CNOT gates along with multiple single-qubit gates~\citep{MikeIke}. 
Optimized for quantum processors with nearest-neighbor connectivity. 
}
    \label{fig:CCX_8_CX}
\end{figure*}

Traditional decomposition methods often assume ideal, fully connected architectures—an unrealistic constraint for most of current quantum hardware. Alternative approaches, such as linear-topology-compatible decompositions using eight CNOT gates, have been explored~\cite{CCX8CNOTS} to address these limitations. A three-qubit CCX gate implementation requires 8 CNOT gates along with multiple single-qubit gates is depicted in Figure~\ref{fig:CCX_8_CX}. Pulse-level optimizations have also been investigated~\citep{Hardware-Conscious_24}, though these can introduce unintended SWAP operations between control qubits, which may not always be desirable.

Recent work by Bowman \textit{et al.}~\citep{Hardware-Conscious} demonstrates hardware-conscious optimizations for the Toffoli gate~\citep{Toffoli} on IBM Quantum devices~\citep{IBMQuantum}, reducing gate infidelity by 18\% and cutting multi-qubit gate requirements through the use of multi-qubit cross-resonance gates~\citep{Hardware-Conscious}. Similarly, Cruz and Murta present shallow decompositions optimized for varying connectivity constraints, employing SWAP-efficient qubit reordering to minimize CNOT overhead~\citep{CruzandMurta}.

Nemkov \textit{et al.} introduce a variational synthesis method that co-optimizes architecture and parameters~\citep{Nemkov}, achieving efficient decompositions for linear and star topologies. Ren \textit{et al.} further advance hardware-aware circuit knitting by integrating gate cuts and SWAP insertions based on qubit interaction graphs~\citep{Ren2024}. Bennakhi \textit{et al.} compare ancilla-based depth reduction techniques for multi-controlled X gates~\citep{Bennakhi}, highlighting the trade-offs between v-chain and recursive methods on NISQ hardware.

Wang \textit{et al.} propose a hardware-efficient quantum random access memory implementation using native {$i$SCZ, C-$i$SCZ} gates, eliminating decomposition overhead while maintaining noise resilience~\citep{Wang}. Duckering \textit{et al.}’s ``Orchestrated Trios" approach~\citep{Duckering} treats Toffoli gates as monolithic units during routing, deferring decomposition to post-mapping stages for improved CNOT efficiency. The study in~\citep{Toffoli_ECR} introduces a hardware-efficient ECR-based~\citep{ibm433,64QV,ECR_Hamiltonian,cross-resonance} Toffoli decomposition for superconducting architectures~\citep{IBMQuantum}. Experimental validation on IBM’s 127-qubit \textit{ibm\_sherbrooke} device demonstrates consistently high fidelity across all operational modes: non-target state preservation ($93.8\% \pm 0.3\%$), target qubit activation ($94.4\% \pm 0.3\%$), and deactivation ($94.3\% \pm 0.3\%$)~\citep{Toffoli_ECR}. This implementation eliminates SWAP requirements in linear topologies, showcasing robust performance under real-world NISQ conditions~\citep{NISQ18}.

\begin{figure*}[htp!]
    \centering
\includegraphics[width=0.9\textwidth]{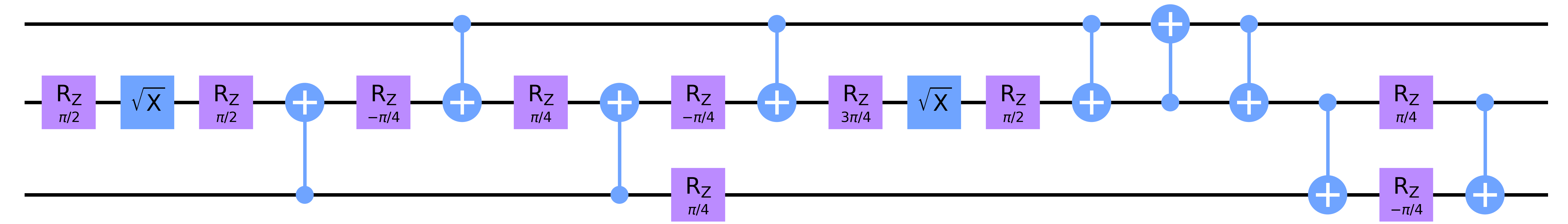}
\caption{
Hardware-aware quantum circuit implementation of the Toffoli (CCNOT) 
gate for linearly connected qubit architectures. The three-qubit operation is decomposed into a sequence of 9 CNOT gates interleaved with single-qubit rotations, compiled for quantum processors with nearest-neighbor connectivity. Our implementation utilizes only the following single-qubit gates: the phase rotation gate $R_z(\varpi)$ for Z-axis rotations, and the $\sqrt{X}$ gate for superposition generation. These gates are represented by the unitary matrices: 
\( 
R_z(\varpi) = \begin{pmatrix} e^{-i\varpi/2} & 0 \\ 0 & e^{i\varpi/2} \end{pmatrix}, \quad 
\sqrt{X} = \frac{1}{2}\begin{pmatrix} 1+i & 1-i \\ 1-i & 1+i \end{pmatrix}.
\)
}
    \label{fig:toffoli_circuit}
\end{figure*}

\begin{figure*}[htp!]
    \centering
\includegraphics[width=0.9\textwidth]{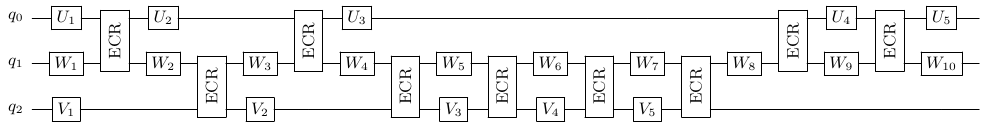}
    \caption{Hardware-aware quantum circuit synthesis of a Toffoli gate (CCNOT) using echoed cross-resonance (ECR) gates and optimized single-qubit gates~\citep{Toffoli_ECR}. The circuit demonstrates native gate decomposition for superconducting quantum processors, with ECR operations constrained to adjacent qubits and single-qubit unitaries ($U_i$, $W_i$, $V_i$) compiled into IBM Quantum's native gate set. 
}
    \label{fig:toffoli_ecr_decomposition}
\end{figure*}

Figure~\ref{fig:toffoli_circuit} introduce a hardware-aware quantum circuit implementation of the Toffoli gate for linearly connected qubit architectures. The CCX gate is decomposed into a sequence of 9 CNOT gates with single-qubit rotations. Additionally, Figure~\ref{fig:toffoli_ecr_decomposition} depicts a hardware-aware Toffoli gate implementation using ECR gates and single-qubit unitaries ($U_i$, $W_i$, $V_i$), tailored for superconducting architectures with nearest-neighbor connectivity~\citep{IBMQuantum}.

\section{Experimental Setup}\label{sec:Methods}

Our experimental protocol for characterizing the Toffoli gate~\citep{Toffoli}  employed a combination of benchmarking techniques (QST and QPT) with comprehensive noise analysis across IBM's superconducting quantum processors~\citep{IBMQuantum}. We employed both the 127-qubit (\textit{ibm\_sherbrooke}) and (\textit{ibm\_brisbane}) quantum architectures. Device calibration parameters, including qubit coherence times (\(T_1\), \(T_2\)), operating frequencies, anharmonicities, and readout fidelities, were meticulously recorded (as detailed in Section~\ref{sec:QuantumHardwarePerformance}). All hardware results were compared against idealized quantum simulations performed using IBM's \textit{qasm\_simulator}, which provided theoretical performance baselines under noiseless conditions.

\subsection{Gate Implementation}\label{sec:Methods_Gate}

Superconducting quantum processors operate in the NISQ era~\citep{SuperconductingQuantumComputers}, where gate fidelities are fundamentally limited by decoherence processes, inter-qubit crosstalk, and calibration drift~\citep{NISQ18}. These challenges are particularly acute for the Toffoli gate~\citep{Toffoli}, where the combination of multi-qubit interactions and extended gate sequences amplifies noise susceptibility. Entangled input states such as GHZ and W states exhibit heightened vulnerability to correlated errors compared to separable states, as their non-local correlations propagate and magnify single-qubit noise channels~\citep{Noise-Assisted}.

\begin{figure*}[htp!]
    \centering
    \includegraphics[width=\textwidth]{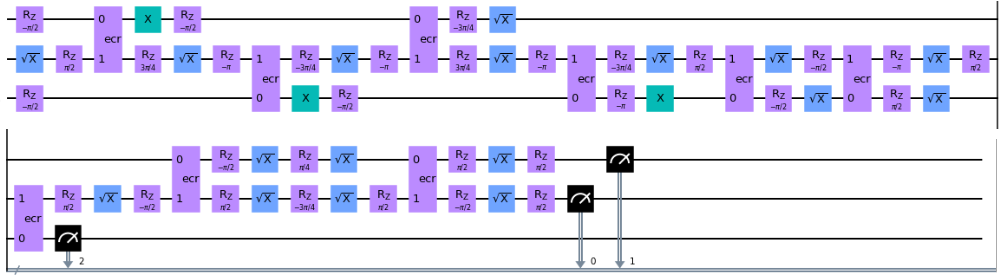}
    \caption{Hardware-aware synthesis of a Toffoli gate using the 2-qubit ECR gates and single-qubit gates. The circuit illustrates a native gate decomposition tailored for superconducting quantum processors, compiled into IBM Quantum's native gate set, \( \mathcal{S}= \{\texttt{ECR, ID, R$_z(\varpi)$, SX, X}\}\), for both the \textit{ibm\_sherbrooke} and \textit{ibm\_brisbane} architectures. Here, $X$ denotes the Pauli-X gate (bit-flip, $|0\rangle \leftrightarrow |1\rangle$), $R_z(\varpi)$ represents a phase rotation around the Z-axis, and $\sqrt{X}$ generates superpositions.  
}
    \label{fig:ECR_IBM}
\end{figure*}

The Toffoli gate~\citep{Toffoli} was implemented using a combination of native gate decompositions. On superconducting hardware, the gate was synthesized into echoed cross-resonance gates and single-qubit rotations, adhering to each processor’s connectivity constraints. The decomposition followed the implementation shown in Figure~\ref{fig:ECR_IBM}.

The implementation of Toffoli gates via echoed cross-resonance protocols introduces additional calibration complexity, requiring precise tuning of microwave pulses to suppress phase errors and mitigate residual ZZ interactions~\citep{ECR_Hamiltonian}. This dual challenge—state-dependent noise sensitivity and exacting pulse-level calibration—underscores the delicate balance between gate functionality and fidelity in current superconducting architectures.

The unitary matrix representations for the two-qubit ECR gates are given by:
\begin{equation} \label{eq:ecr1}
ECR_{q_0\rightarrow q_1} = 
\begin{pmatrix} 
0 & 1/\sqrt{2} & 0 & i/\sqrt{2} \\ 
1/\sqrt{2} & 0 & -i/\sqrt{2} & 0 \\ 
0 & i/\sqrt{2} & 0 & 1/\sqrt{2} \\ 
-i/\sqrt{2} & 0 & 1/\sqrt{2} & 0 
\end{pmatrix}
\end{equation}
\begin{equation} \label{eq:ecr2}
ECR_{q_1\rightarrow q_0} = 
\begin{pmatrix} 
0 & 0 & 1/\sqrt{2} & i/\sqrt{2} \\ 
0 & 0 & i/\sqrt{2} & 1/\sqrt{2} \\ 
1/\sqrt{2} & -i/\sqrt{2} & 0 & 0 \\ 
-i/\sqrt{2} & 1/\sqrt{2} & 0 & 0 
\end{pmatrix}
\end{equation}

\noindent
where $ECR_{q_i\rightarrow q_j}$ implements the echoed cross-resonance interaction with $q_i$ acting as the control and $q_j$ as the target. Our implementation utilizes three native single-qubit gates on the \textit{ibm\_sherbrooke} superconducting processor 
(Section~\ref{sec:QuantumHardwarePerformance}): the Pauli-X gate ($X$) for bit-flip operations ($|0\rangle \leftrightarrow |1\rangle$), the phase rotation gate $R_z(\varpi)$ for Z-axis rotations, and the $\sqrt{X}$ gate for superposition generation. These gates are represented by the unitary matrices:

\begin{equation}
X = \begin{pmatrix} 0 & 1 \\ 1 & 0 \end{pmatrix}
\end{equation}

\begin{equation}
R_z(\varpi) = \begin{pmatrix} e^{-i\varpi/2} & 0 \\ 0 & e^{i\varpi/2} \end{pmatrix}
\end{equation}

\begin{equation}
\sqrt{X} = \frac{1}{2}\begin{pmatrix} 1+i & 1-i \\ 1-i & 1+i \end{pmatrix}.
\end{equation}

The $R_z(\varpi)$ gate preserves computational basis probabilities while applying phase shifts, whereas $\sqrt{X}$ serves as the principal square root of $X$ ($(\sqrt{X})^2 = X$), enabling coherent transitions between basis states. Together with two-qubit entangling gates, this set 
(\( \mathcal{S}_{\text{sherbrooke}}= \{\texttt{ECR, ID, R$_z(\varpi)$, $\sqrt{\text{X}}$, X}\}\)) 
provides universal quantum computation capabilities on the \textit{ibm\_sherbrooke} quantum architecture.

\subsection{State Preparation}\label{sec:method_state}

To evaluate the gate’s performance across distinct operational regimes, we prepared three distinct classes of input states representing different entanglement configurations: 
\begin{enumerate}
     \item The maximally entangled GHZ state: \(|\psi_{\text{GHZ}}\rangle = 1/\sqrt{2}(|000\rangle + |111\rangle)\), which test coherence in maximally entangled systems. 
     \item The W state: \(|\psi_{\text{W}}\rangle = \frac{1}{\sqrt{3}}(|001\rangle + |010\rangle + |100\rangle)\), sensitive to asymmetric error. 
     \item A uniform superposition of all three-qubit computational basis states: \( |\psi_{\text{unif}}\rangle = \frac{1}{\sqrt{8}}\sum_{i=0}^7 |b_i\rangle\), where \(b_i\) spans \(|000\rangle\) to \(|111\rangle\),  to assess gate performance under balanced computational basis conditions. 
\end{enumerate}

\subsection{Tomography Protocol}\label{sec:method_tomography}

QPT and QST are fundamental characterization techniques in quantum computing used to analyze quantum systems. QST is a procedure for reconstructing a quantum system's state from experimental measurements. The QST protocol involves:
repeated measurements in different bases; state preparation before each measurement round; and computational reconstruction of the density matrix $\rho$ from measurement statistics. The technique requires careful calibration as the number of measurements grows exponentially with system size. Various reconstruction methods have been developed, including maximum likelihood estimation and compressed sensing approaches~\citep{QPTGambetta,QST06,QST08,QST09,QST10,QST11,QST_epj}.

In our experiments, QST was employed to reconstruct the output density matrix (\(\rho_{\text{out}}\)) for each input state. Measurements were performed in the Pauli bases across all qubits, for each basis configuration, we collected 19,000 shots for \(|\psi_{\text{GHZ}}\rangle\) and \(|\psi_{\text{W}}\rangle\) states, and 11,000 shots for the uniform superposition \( |\psi_{\text{unif}}\rangle\) to ensure statistical robustness. The state fidelity $\mathcal{F}_S(\varrho, \pounds)$ of each output state~\citep{Fidelity} was computed as shown in Equation~\eqref{eq:state} and uncertainty estimates were derived.

QPT is a key method for fully characterizing quantum operations by analyzing input-output state correlations~\citep{MikeIke,chuang97,brien,qpt}. It reconstructs a process matrix that captures all possible quantum state transitions, enabling precise benchmarking of gate fidelities and error identification~\citep{Bialczak,Entangling}. Essential for validating quantum hardware and algorithms, QPT provides critical metrics for scaling reliable quantum computers~\citep{AGSycam1,AGSycam2}.

QPT characterizes quantum operations (gates or channels) through experimental benchmarking. Key aspects include:
application to a complete set of input states; measurement of corresponding outputs; and reconstruction of the process matrix $\chi$ or Choi matrix~\citep{Choi}. QPT is particularly valuable for: gate validation in quantum processors~\citep{AGSycam1,AGSycam2}; noise characterization in quantum circuits and calibration of experimental setups~\cite{SQSCZ1}. Theoretical foundations derive from the operator-sum representation~\citep{kraus83,chuang97} and quantum channel theory~\citep{qpt,chell03,brien}. Modern implementations often use the Choi-Jamiołkowski isomorphism~\citep{Choi,Jamiołkowski} for complete process characterization~\cite{SQSCZ2,Entangling}.

QPT works by preparing a complete set of input states and performing measurements in various bases to reconstruct an unknown quantum process. In Qiskit's implementation~\cite{qiskit}, each qubit is initialized in one of four probe states spanning the Bloch sphere: $\{\ket{0}, \ket{1}, \ket{\psi_+}, \ket{\phi_+}\}$, corresponding to the eigenstates of the Pauli operators $Z$, $X$, and $Y$, respectively. Measurements are performed in each of the three Pauli bases ($X$, $Y$, and $Z$), using projectors defined as:

\begin{equation}
\begin{aligned}
P_{x+} &= \ket{\psi_+}\bra{\psi_+}, &\quad P_{z+} &= \ket{0}\bra{0}, \\
P_{y+} &= \ket{\phi_+}\bra{\phi_+}, &\quad P_{z-} &= \ket{1}\bra{1},
\end{aligned}
\end{equation}

where the superposition states are given by:
\begin{equation}
\begin{aligned}  
\ket{\psi_\pm} &= \frac{1}{\sqrt{2}}(\ket{0} \pm \ket{1}), \quad  
\ket{\phi_\pm} &= \frac{1}{\sqrt{2}}(\ket{0} \pm i\ket{1}).
\end{aligned}
\end{equation}

For a $k$-qubit system, quantum process tomography involves measurements in all $3^k$ combinations of single-qubit Pauli bases, i.e., $\{X, Y, Z\}^{\otimes k}$.
To fully characterize a $k$-qubit quantum process, this procedure requires executing $12^k$ distinct quantum circuits: $4^k$ linearly independent input states combined with $3^k$ measurement settings. Each circuit must be run multiple times on actual quantum hardware to collect sufficient statistics for reliable process matrix reconstruction. This exponential scaling presents a significant resource challenge for QPT on systems with more than a few qubits, particularly on real hardware with limited shot budgets and coherence times.

The measurement results are used to construct a Choi matrix that mathematically represents the reconstructed quantum channel~\citep{Choi,Jamiołkowski}. The average gate fidelity \( F_{\text{ave}}(\Im, \mathcal{W}) \) is then calculated, a single number between 0 and 1, by comparing this reconstructed channel with the ideal target unitary operation~\citep{Horodecki,Nielsen02}. A fidelity value closer to 1 indicates that the implemented quantum process more accurately matches the desired unitary operation,, with deviations from the simulator attributed to hardware-specific noise. This fidelity metric serves as a valuable tool for comparing different implementations of quantum gates~\citep{Horodecki,Nielsen02}.

\subsection{Error Analysis}

The average gate fidelity \( F_{\text{ave}}(\Im, \mathcal{W}) \) provides a quantitative measure of how accurately a physical quantum channel \( \Im \) implements a desired unitary operation \( \mathcal{W} \)~\citep{Horodecki,Nielsen02}. This metric admits two mathematically equivalent expressions:

\begin{equation}
    \mathcal{F}_{\text{ave}}(\Im, \mathcal{W}) = \int \mathrm{d}\psi\, \langle \psi | \mathcal{W}^\dagger \Im(|\psi\rangle\langle\psi|) \mathcal{W} | \psi \rangle
\end{equation}

and alternatively as:

\begin{equation}
    \mathcal{F}_{\text{ave}}(\Im, \mathcal{W}) = \frac{\Gamma \cdot \mathcal{F}_{\text{pro}}(\Im, \mathcal{W}) + 1}{\Gamma + 1}
\end{equation}

\noindent
Where, \( \Gamma = 2^k \) denotes the dimension of the Hilbert space for a \( k \)-qubit system, and \( \mathcal{F}_{\text{pro}}(\Im, \mathcal{W}) \) represents the process fidelity between the channel and target unitary. 

The process fidelity \( \mathcal{F}_{\text{pro}}(\Theta, \Psi) \) between arbitrary quantum channels \( \Theta \) and \( \Psi \) is defined through their Choi representations:

\begin{equation}
    \mathcal{F}_{\text{pro}}(\Theta, \Psi) = \mathcal{F}_S(\sigma_\Theta, \sigma_\Psi)
\end{equation}
\noindent
where:
\( \sigma_\Theta = \Xi_\Theta/\Gamma \) is the normalized Choi matrix of channel \( \Theta \), 
\( \sigma_\Psi = \Xi_\Psi/\Gamma \) is the normalized Choi matrix of channel \( \Psi \), and 
$\mathcal{F}_{\text{sta}}$ denotes the standard state fidelity between density operators~\citep{Fidelity}.

The underlying quantum state fidelity between density operators \( \varrho \) and \( \pounds \) is given by:

\begin{equation}\label{eq:state}
    \mathcal{F}_S(\varrho, \pounds) = \left( \text{Tr} \sqrt{ \sqrt{\varrho} \pounds \sqrt{\varrho} } \right)^2
\end{equation}

\noindent
Thus, $\mathcal{F}_{\text{pro}}(\Theta, \Psi)$  can be written as: 

\begin{equation}
    \mathcal{F}_{\text{pro}}(\Theta, \Psi) = \left( \mathrm{Tr} \sqrt{ \sqrt{\sigma_\Theta}\, \sigma_\Psi \sqrt{\sigma_\Theta} } \right)^2
\end{equation}

When comparing a channel to a unitary operation \( \mathcal{W} \), this reduces to:
\begin{equation}
    \mathcal{F}_{\text{pro}}(\Im, \mathcal{W}) = \frac{ \mathrm{Tr}(S_\mathcal{W}^\dagger S_{\Im}) }{\Gamma^2}
\end{equation}

\noindent
where \( S_{\Im} \) and \( S_\mathcal{W} \) denote the superoperator representations in the Pauli basis. If the target channel \(\Psi\) has a pure state 
\(\sigma_\Psi = |\psi_\Psi\rangle\langle\psi_\Psi|\), the fidelity simplifies to:
\begin{equation}
\mathcal{F}_{\text{pro}}(\Theta, \Psi) = \langle \psi_\Psi | \sigma_\Theta | \psi_\Psi \rangle
\end{equation}

\section{Results and Discussion} \label{SEC:results}

\begin{figure*}[htp!]
    \centering
    \includegraphics[width=0.75\textwidth]{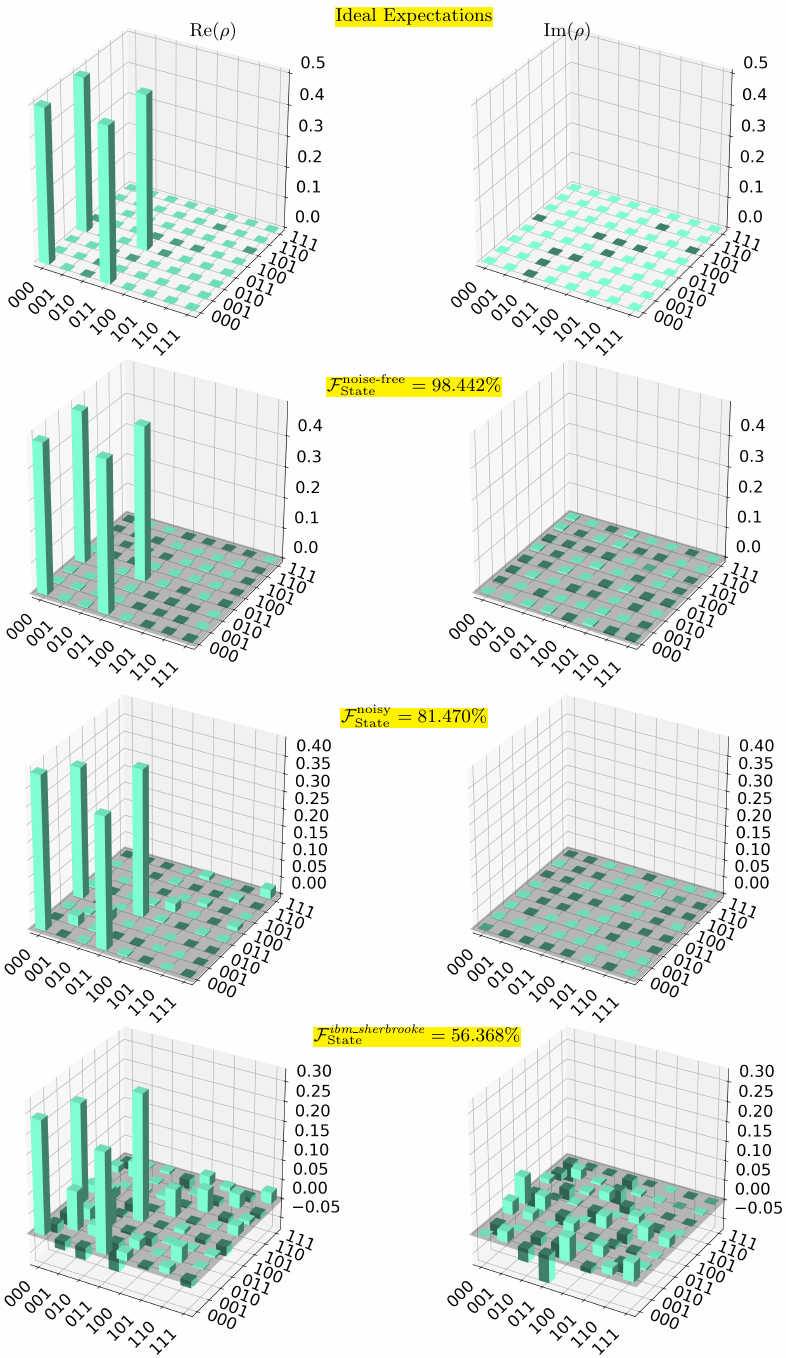}
    \caption{QST of the Toffoli gate with input state prepared as the maximally entangled GHZ state (\( \frac{1}{\sqrt{2}}(|000\rangle + |111\rangle)\)). The QST measurements were conducted with 19,000 shots in each different operational environment. The results demonstrate substantial variations in gate fidelity across different operational environments, achieving fidelities of 98.442\% (noise-free), 81.470\% (noisy emulation), and 56.368\% (real hardware, \textit{ibm\_sherbrooke}) with hardware implementations exhibiting particular sensitivity to initial state preparation conditions. 
    }
    \label{fig:qst_ghz}
\end{figure*}

\begin{figure*}[htp!]
    \centering
    \includegraphics[width=0.75\textwidth]{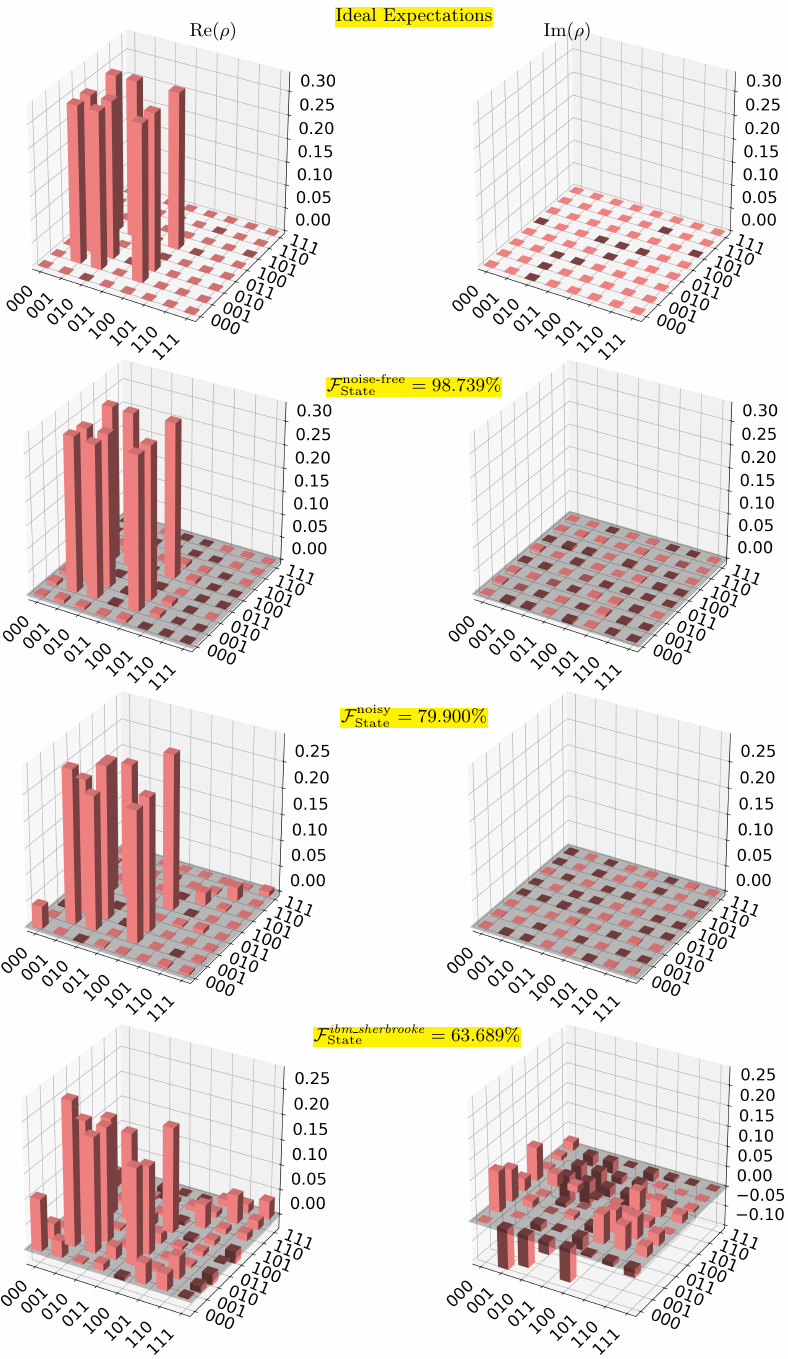}
    \caption{QST of the Toffoli gate with input state prepared as W states (\( \frac{1}{\sqrt{3}} (|001\rangle + |010\rangle + |100\rangle)\)). The QST measurements were conducted with 19,000 shots in each different operational environment. The results demonstrate substantial variations in gate fidelity across different operational environments, achieving fidelities of 98.739\% (noise-free), 79.900\% (noisy emulation), and 63.689\% (real hardware, \textit{ibm\_sherbrooke}) with hardware implementations exhibiting particular sensitivity to initial state preparation conditions. 
    }
    \label{fig:qst_w}
\end{figure*}

\begin{figure*}[htp!]
    \centering
     \includegraphics[width=\textwidth]{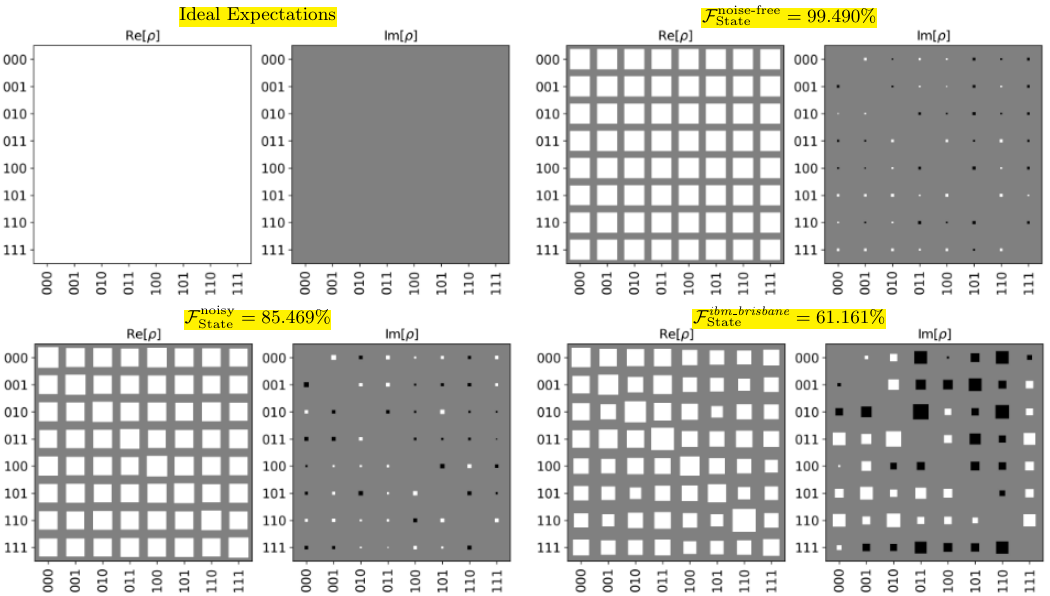}
    \caption{QST of the Toffoli gate with input state prepared as a uniform superposition of all three-qubit computational basis states (\(\frac{1}{\sqrt{8}}\sum_{i=0}^7 |b_i\rangle\), where \(b_i\) spans \(|000\rangle\) to \(|111\rangle\)). The QST measurements were conducted with 11,000 shots in each different operational environment. The results demonstrate substantial variations in gate fidelity across different operational environments, achieving fidelities of 99.490\% (noise-free), 85.469\% (noisy emulation), and 61.161\% (real hardware, \textit{ibm\_brisbane}) with hardware implementations exhibiting particular sensitivity to initial state preparation conditions. 
    }
    \label{fig:qst_unif}
\end{figure*}

\begin{figure*}[htp!]
    \centering
    \includegraphics[width=0.9\textwidth]{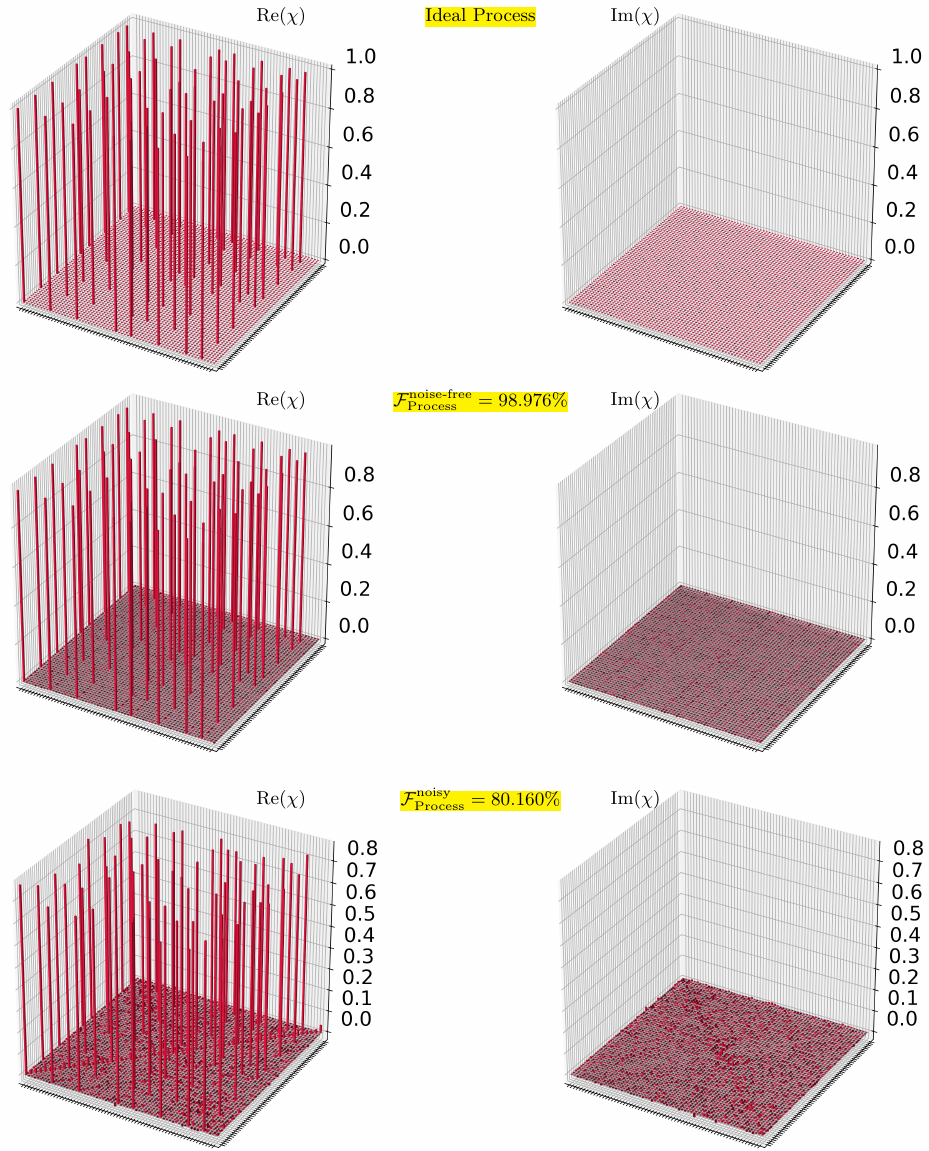}
    \caption{QPT experiments of the Toffoli gate. While the \textit{qasm\_simulator} showed near-ideal process fidelity (99.0\%), noise-aware quantum emulation revealed how quickly performance degrades (80.2\% process fidelity).}
    \label{fig:qpt}
\end{figure*}

The fidelities from QST experiments represent how closely the actual output state of our Toffoli gate matches the ideal expected state for specific input classes (GHZ, W, uniform superposition). Our QST experiments reveal critical insights into the performance of the Toffoli gate~\citep{Toffoli} across different input states and hardware platforms. The GHZ state—a critical benchmark for multi-qubit entanglement—showed perfect behavior in noise-free quantum simulation (98.4\% state fidelity), with slight degradation in noisy quantum emulation (81.470\% state fidelity), but suffered dramatic degradation on hardware. When executed on \textit{ibm\_sherbrooke}, the same gate operation yielded just 56.4\% state fidelity (Figure~\ref{fig:qst_ghz}).

The W state followed a similar pattern, maintaining 98.7\% state fidelity in noise-free quantum simulation,  
79.900\% state fidelity in noisy environment, and dropping to 63.7\% on quantum hardware (Figure~\ref{fig:qst_w}), with its complex superposition particularly vulnerable to asymmetric noise. Interestingly, the uniform superposition state demonstrated the highest noise-free quantum simulation state fidelity (99.5\%) and 85.469\% state fidelity in noisy settings, but still faced substantial challenges on \textit{ibm\_brisbane}, achieving only 61.2\% state fidelity (Figure~\ref{fig:qst_unif}). This universal performance drop across all test states highlights fundamental limitations in current NISQ-era hardware.

QPT quantified the gate's operational limitations with even starker results. While the \textit{qasm\_simulator} showed near-ideal process fidelity (99.0\%), noisy emulation revealed how quickly performance degrades (80.2\% process fidelity). The missing hardware process fidelity measurements would likely show even more severe degradation, consistent with our state tomography results. Figure~\ref{fig:qpt} presents results from our QPT characterization of the Toffoli gate, including the reconstructed quantum process matrices along with their corresponding process fidelities. The protocol requires $12^k$ distinct circuit configurations for a $k$-qubit operation. For the Toffoli gate ($k=3$), we executed each of the $12^3$ circuits with 11,000 measurement shots, resulting in a total of $1.9008 \times 10^7$ individual measurements to ensure statistically reliable results.

The complete QPT of the 3-qubit Toffoli gate~\citep{Toffoli}—requiring 1,728 quantum circuits and 19,008,000 individual measurements—was conducted using both the noise-free \textit{qasm\_simulator} and a noisy quantum emulator, as real hardware execution was infeasible due to experimental constraints. The noisy emulator replicates real quantum hardware by integrating experimentally measured noise profiles, including gate errors, readout errors, and decoherence times. This approach provides a realistic approximation of how QPT behaves on actual IBM quantum processors under realistic noise conditions. By leveraging this emulation framework, we obtain rigorous yet experimentally tractable insights into gate fidelity while avoiding the prohibitive resource overhead of performing full QPT on physical quantum devices.

Table~\ref{table:FID2} presents an analysis of Toffoli gate state-dependent fidelity metrics across different operational environments on IBM's 127-qubit \textit{ibm\_sherbrooke} and \textit{ibm\_brisbane} quantum processors. Through rigorous QST and QPT protocols, we evaluated gate performance for three distinct input state classes: GHZ states, W states, and uniform superpositions. The QST measurements were conducted with 19,000 shots for GHZ and W state analyses, while the uniform superposition characterization employed 11,000 shots to ensure statistically robust results. Corresponding QPT investigations were conducted with 11,000 shots for reliable process fidelity estimation. The results demonstrate substantial variations in gate fidelity across different input states and operational environments, with hardware implementations exhibiting particular sensitivity to initial state preparation conditions, highlighting the considerable implementation challenges in current NISQ processors.

\begin{table*}[!htp]
\caption{\label{table:FID2} Fidelity metrics obtained form a systematic characterization of the Toffoli gate on IBM's 127-qubit \textit{ibm\_sherbrooke} and \textit{ibm\_brisbane} processors using quantum state tomography (QST) and quantum process tomography (QPT). We measure the gate’s fidelity for three input state classes (GHZ, W, and uniform superpositions). All QST measurements used 19,000 shots except uniform superposition (11,000 shots). QPT experiments employed 11,000 shots. Our results demonstrate significant fidelity variations across different input states and environments, with hardware implementations showing particular sensitivity to state preparation. The process fidelity maintains 0.990 in simulation but drops to 0.802 under noisy conditions, revealing substantial implementation challenges for three-qubit gates on current superconducting quantum processors.}
\centering 
\renewcommand{\arraystretch}{1.1}
\begin{tabular}{|l|  c|  c| l|}
\hline \hline
\multirow{2}{*}{Input State} & \multicolumn{3}{c|}{QST experiments} \\
\cline{2-4}
 & Noise-free settings\footnotemark[1] & Noise-aware quantum emulation & Real quantum hardware \\
\hline \hline
GHZ ($\frac{1}{\sqrt{2}}(|000\rangle + |111\rangle)$)             & 98.442\% & 81.470\% & 56.368\% (\textit{ibm\_sherbrooke})\\
W ($\frac{1}{\sqrt{3}}(|001\rangle + |010\rangle + |100\rangle)$) & 98.739\% & 79.900\% & 63.689\% (\textit{ibm\_sherbrooke})\\
Uniform superpositions 
($\frac{1}{\sqrt{8}}\sum_{x \in \{0,1\}^3}|x\rangle$)             & 99.490\% & 85.469\% & 61.161\% (\textit{ibm\_brisbane})\\
\hline
QPT (Process Fidelity) & 99.0 \%& 80.2\% & --\footnotemark[2] \\
\hline \hline
\end{tabular}
\footnotetext[1]{“Noise-free” refers to an idealized quantum circuit simulated with finite-shot sampling, 
hence the fidelity $<100\%$ is due to statistical (sampling) error, not physical noise.}
\footnotetext[2]{Full QPT of the Toffoli gate—requiring 1,728 circuits—was executed using both noise-free simulation and noise-aware emulation, as real hardware execution was constrained by runtime and coherence limitations.
}
\end{table*}

\section{Quantum Hardware Performance}
\label{sec:QuantumHardwarePerformance}

\begin{figure*}[htp!]
    \centering
    \includegraphics[width=0.95\textwidth]{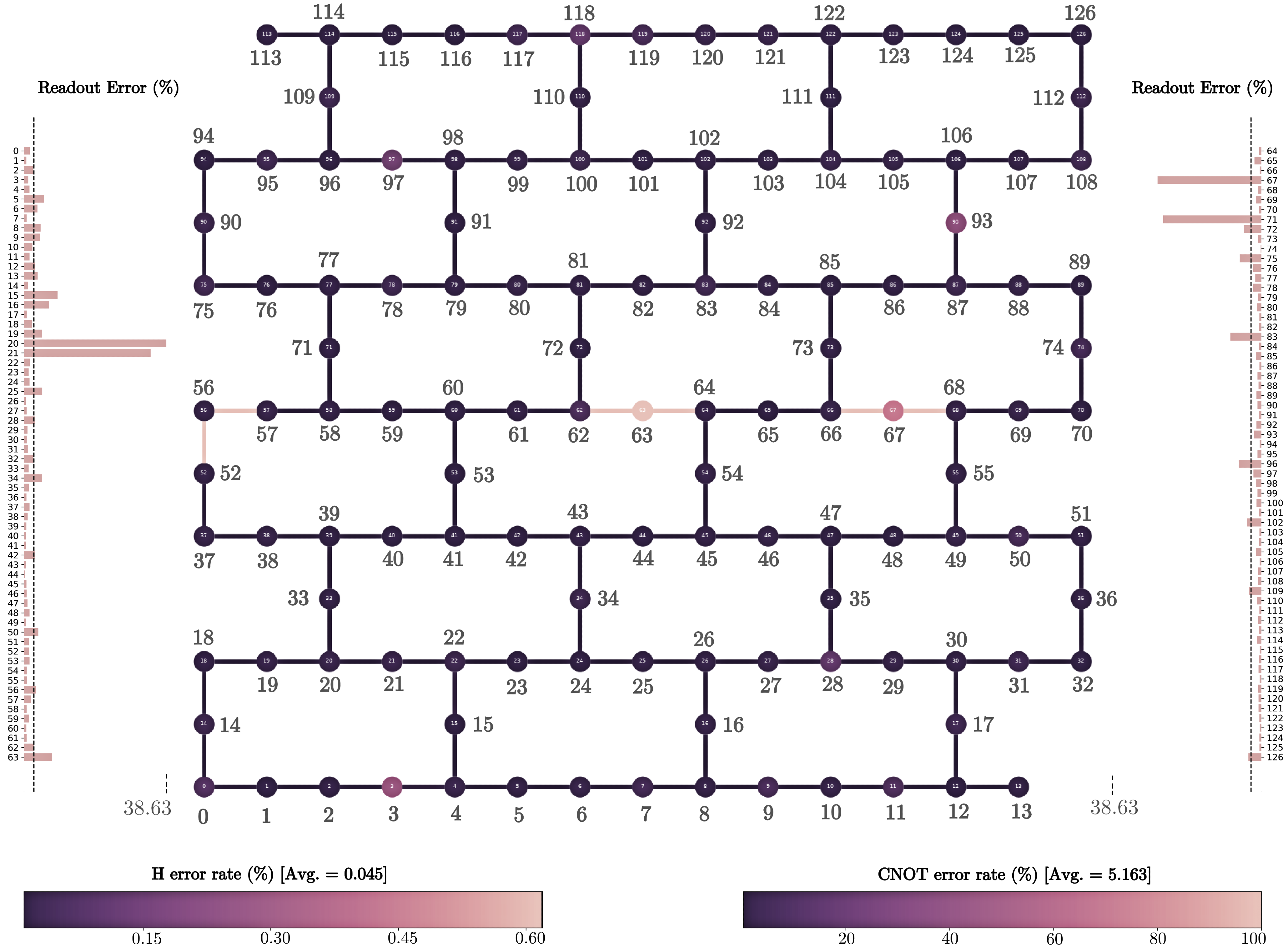}
    \caption{
Qubit connectivity and readout error map of the \textit{ibm\_sherbrooke} quantum computer, featuring a heavy-hexagonal architecture where each qubit (circles) connects to 2-3 neighbors.  With qubit-specific readout error rates, demonstrating performance variations across the device. The layout reflects the architecture of IBM's \textit{Eagle} r3 processor, identical layout implemented in the \textit{ibm\_brisbane} quantum system. This visualization was regenerated from~\citep{GSA} under a Creative Commons Attribution 4.0 International License (\url{https://creativecommons.org/licenses/by/4.0/}). 
   }
    \label{fig:sherbrooke}
\end{figure*}

\begin{table*}[!htp]
\caption{
    Statistical summary of qubit performance metrics for \textit{ibm\_sherbrooke} during Toffoli gate characterization using GHZ states. Data obtained through QST with input state $|\psi_{\text{GHZ}}\rangle = \frac{1}{\sqrt{2}}(|000\rangle + |111\rangle)$. 
    Metrics were recorded during QST execution to ensure operational relevance. Measured parameters, 
including coherence times ($T_1$, $T_2$), 
qubit frequencies ($\omega/2\pi$), anharmonicities ($\alpha/2\pi$), 
probability of misclassifying $|1\rangle$ as $|0\rangle$ ($P(\ket{0}|\ket{1})$), 
probability of misclassifying $|0\rangle$ as $|1\rangle$ ($P(\ket{1}|\ket{0})$), 
readout errors ($\epsilon_r$), and readout length ($\tau_r$). Metrics were recorded during QST execution to ensure operational relevance.
The table presents the mean, standard deviation (std), minimum (min), quartiles (25\%, 50\%, 75\%), and maximum (max) values for each measured parameter across the qubit ensemble. All values are reported with their original units and precision. 
}
    \label{tab:performancemetrics_GHZ}
    \begin{tabular}{c}
    \includegraphics[width=0.8\textwidth]{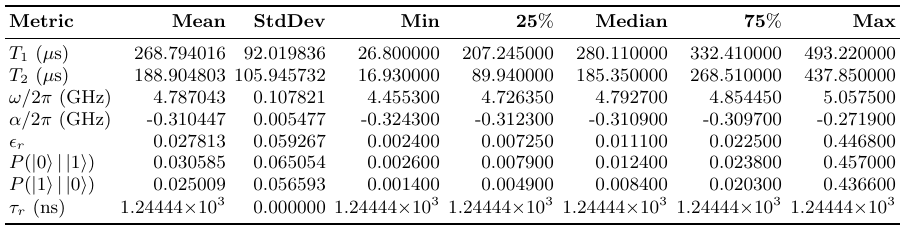} \\
    \end{tabular}
\end{table*}

\begin{table*}[!htp]
\caption{
    Statistical summary of qubit performance metrics for \textit{ibm\_sherbrooke} during Toffoli gate characterization using W states. Data obtained through QST with input state $|\psi_{\text{W}}\rangle = \frac{1}{\sqrt{3}}(|001\rangle + |010\rangle + |100\rangle)$. 
}
    \label{tab:performancemetrics_W} 
    \begin{tabular}{c}
    \includegraphics[width=0.8\textwidth]{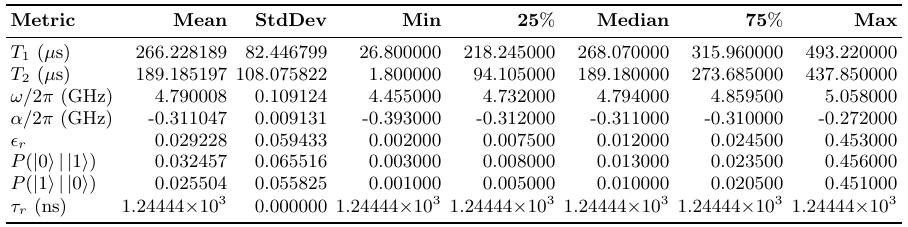} \\
    \end{tabular}
\end{table*}

\begin{table*}[!htp]
\caption{
Statistical summary of performance metrics for \textit{ibm\_brisbane} recorded during QST execution to ensure operational relevance. Data obtained through QST for Toffoli gate with uniform superpositions: with input state $|\psi_{\text{unif}}\rangle = \frac{1}{\sqrt{8}}\sum_{x=0}^7 |x\rangle$, where $|x\rangle$ represents 3-bit computational basis states. 
}
    \label{tab:performancemetrics_Unif}
    \begin{tabular}{c}
    \includegraphics[width=0.8\textwidth]{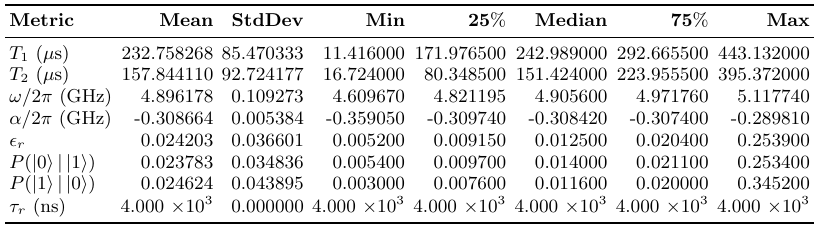} \\
    \end{tabular}
\end{table*}

Our experiments were conducted on two IBM's \textit{Eagle} r3 superconducting quantum processors: \textit{ibm\_sherbrooke}, \textit{ibm\_brisbane}. Both state-of-the-art 127-qubit quantum processors implement the ECR gates as their native entangling operations, with median error rates of $7.56 \times 10^{-3}$ (\textit{ibm\_sherbrooke}) and $8.32 \times 10^{-3}$ (\textit{ibm\_brisbane}) respectively. Isolated two-qubit gate fidelities reached $2.92 \times 10^{-3}$ (\textit{ibm\_sherbrooke}) and $3.16 \times 10^{-3}$ (\textit{ibm\_brisbane}) as measured by randomized benchmarking, while layered error per logical gate (EPLG) for 100-qubit chains was $1.35 \times 10^{-2}$ and $1.65 \times 10^{-2}$ respectively. The systems demonstrated distinct performance characteristics: \textit{ibm\_sherbrooke}  exhibited median coherence times ($T_1 = 272.21 ~\mu s, T_2 = 188.10 ~\mu$s) and lower gate errors, making it preferable for deep circuits, whereas \textit{ibm\_brisbane} achieved higher computational throughput ($180$K vs $150$K CLOPS), benefiting real-time applications. Both systems shared identical basis gate sets ($\mathcal{S} = \texttt{\{ECR, ID, R$_z(\varpi)$, $\sqrt{\text{X}}$, X\}}$) and support pulse-level control capabilities, with median single-qubit ($\sqrt{\text{X}}$) and readout errors below $2.36\times 10^{-4}$ and $1.71 \times 10^{-2}$ respectively. This measured performance characteristics (as of 14 April 2025)  represents the state-of-the-art performance for NISQ-era superconducting quantum processors.

The Toffoli gate~\citep{Toffoli} implementations were characterized across three distinct experimental configurations, with comprehensive qubit metrics documented in Tables s~\ref{tab:performancemetrics_GHZ}--\ref{tab:performancemetrics_Unif}. 
Tables~\ref{tab:performancemetrics_GHZ} and ~\ref{tab:performancemetrics_W} detail \textit{ibm\_sherbrooke}'s behavior for entangled input states, with Table~\ref{tab:performancemetrics_GHZ} quantifying GHZ-state operations ($|\psi_{\text{GHZ}}\rangle$), 
and Table~\ref{tab:performancemetrics_W} capturing W-state dynamics ($|\psi_{\text{W}}\rangle$). Table~\ref{tab:performancemetrics_Unif}  presents \textit{ibm\_brisbane}'s performance during uniform superposition tests ($|\psi_{\text{unif}}\rangle$), revealing median coherence times under balanced state preparation of $T_1$ = 242.989 $~\mu$s and $T_2$ = 151.424 $~\mu$s, with isolated ECR gate errors reaching $3.16\times 10^{-3}$. This dataset enables comparative analysis of error propagation across different entanglement regimes while maintaining consistent operational conditions.

\begin{figure*}[htp!]
    \centering
    \includegraphics[width=0.9\textwidth]{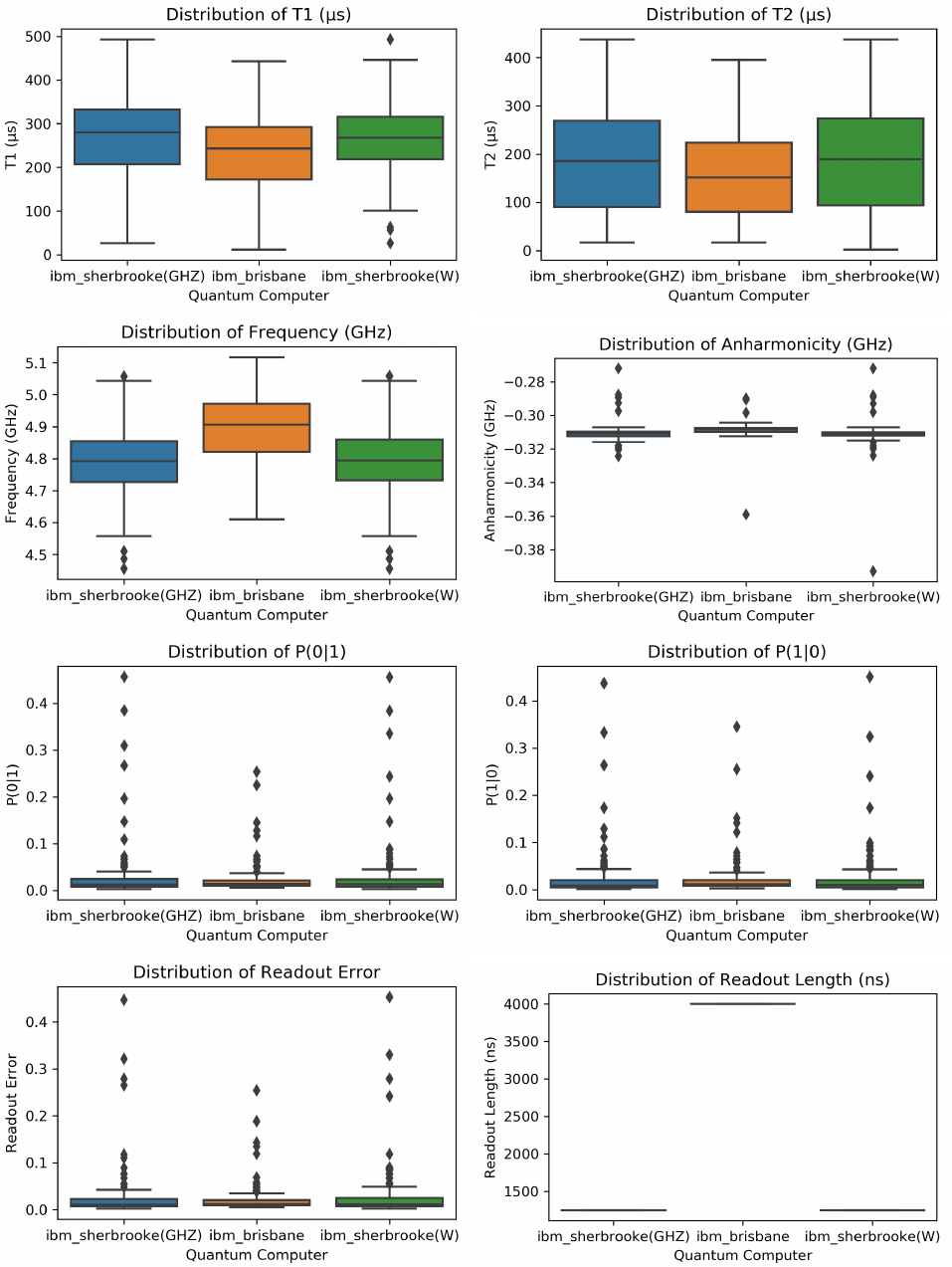}
    \caption{
        Comparative distributions of qubit performance metrics across the quantum processors. Boxplots show measured values (rows from upper left to lower right) of: 
        coherence times ($T_1$, $T_2$ in $\mu$s), 
        transition frequencies ($f_{01}$) and anharmonicities ($\alpha$, both in GHz), 
        readout error probabilities ($P(0|1)$, $P(1|0)$), and 
        readout errors and pulse lengths (ns). Data were collected during Toffoli gate characterization using three distinct input states: GHZ states ($|\psi_\text{GHZ}\rangle = \frac{1}{\sqrt{2}}(|000\rangle + |111\rangle)$) on \textit{ibm\_sherbrooke}, W states ($|\psi_\text{W}\rangle = \frac{1}{\sqrt{3}}(|001\rangle + |010\rangle + |100\rangle)$) on \textit{ibm\_sherbrooke}, and uniform superpositions ($|\psi_\text{unif}\rangle = \frac{1}{\sqrt{8}}\sum_{x=0}^7|x\rangle$) on \textit{ibm\_brisbane}. Whiskers indicate 5th-95th percentiles of the distributions across all 127 qubits.}
    \label{fig:Boxplots}
\end{figure*}

Figure~\ref{fig:sherbrooke} shows the qubit connectivity and readout errors of IBM's 127-qubit \textit{ibm\_sherbrooke} processor (\textit{Eagle} r3 architecture), which shares the same layout as \textit{ibm\_brisbane}. Figure~\ref{fig:Boxplots} presents a comparison 
of qubit performance metrics across three quantum processor configurations, measured during Toffoli gate characterization. Data were collected under three distinct input states: GHZ states ($|\psi_\text{GHZ}\rangle$) 
and W states ($|\psi_\text{W}\rangle $)
on \textit{ibm\_sherbrooke}, and uniform superpositions ($|\psi_\text{unif}\rangle $) 
on \textit{ibm\_brisbane}. Whiskers indicate the 5th–95th percentile ranges across all 127 qubits, demonstrating state-dependent performance characteristics in superconducting quantum hardware.

\newpage

\section{Conclusion}\label{SEC:conclusion}

This study presents a rigorous experimental characterization  of the Toffoli gate’s performance on state-of-the-art superconducting quantum processors, combining QST and QPT to uncover the intricate relationship between input states, gate fidelity, and hardware-specific noise, 
revealing fundamental insights into the challenges of executing three-qubit operations in the NISQ era.     
We evaluate the gate performance across entangled and separable input states using QST and QPT. 
We evaluate the gate’s fidelity for three classes of input states: Greenberger–Horne–Zeilinger states ($|\psi_\text{GHZ}\rangle$), achieving fidelities of 98.442\% (noise-free quantum simulation), 81.470\% (noise-aware quantum emulation), and 56.368\% (real quantum hardware);  W states ($|\psi_\text{W}\rangle$), yielding state fidelities of 98.739\%, 79.900\%, and 63.689\% under the same conditions, while a  uniform superposition of all three-qubit computational basis states ($|\psi_\text{unif}\rangle$), reaching 99.490\%, 85.469\%, and 61.161\%, respectively.

The state-dependent fidelity patterns we observed are particularly revealing. Our quantum tomography-based approach demonstrates that while noise-free quantum simulations predict near-perfect operation ($98.4-99.5\%$ state fidelity), and 
noise-aware quantum emulation predict $79.90-85.469\%$ state fidelity; actual hardware performance remains severely limited, achieving $56-64\%$ state fidelity across different input state classes. Additionally, QPT experiments reveal a process fidelity of 98.976\% and 80.160\% in noise-free and noise-aware quantum emulation, respectively. This precipitous decline underscores the substantial challenges facing NISQ-era quantum computing.

The observed performance variations across different input states—particularly the pronounced state fidelity degradation in entangled states compared to simpler superpositions—highlight the complex interplay between quantum state preparation and gate implementation. The demonstrated state-dependent performance characteristics suggest that algorithm designers have to carefully consider input state selection and may need to develop specialized error mitigation strategies for operations involving three-qubit gates.

These findings provide valuable insights into the operational limitations of complex quantum gates on near-term quantum hardware.  
As quantum processors scale, the insights gained here will inform the development of robust, application-specific gate implementations—bridging the gap between theoretical potential and practical execution in the NISQ era and beyond.

\begin{acknowledgments}

``We acknowledge the use of IBM Quantum's computers for this research. However, the views and conclusions expressed by the author are their own and do not necessarily reflect IBM Quantum’s official policy or position."  

\end{acknowledgments}

\section*{Declarations}

\subsection*{Ethical approval and consent to participate}
Not applicable.

\subsection*{Consent for publication}
The author have approved the publication. This research did not involve any human, animal or other participants.

\subsection*{Availability of supporting data}
The datasets generated during and/or analyzed during the current study are included within this article.

\subsection*{Competing interests}
The author declares no competing interests.

\subsection*{Funding}

This article is published open access under the agreement between Springer Nature and the Science, Technology \& Innovation Funding Authority (STDF), in cooperation with the Egyptian Knowledge Bank (EKB).
The author declares that no funding, grants, or other forms of financial support were received at any stage of this research.

\subsection*{Authors' contributions}

``M. AbuGhanem: conceptualization, methodology, resources, quantum programming, experimental implementations on IBM Quantum's quantum computers, data curation, formal analysis, statistical analysis, visualization, investigation, validation, writing, reviewing and editing. The author has approved the final manuscript."

\end{document}